# Fabrication of nanocrystal superlattice microchannels by soft-lithography for electronic measurements of single-crystalline domains

Andre Maier[1,2]*, Ronny Löffler[2] and Marcus Scheele[1,2]

[1] Institute of Physical and Theoretical Chemistry, University of Tübingen, Auf der Morgenstelle 18, D-72076 Tübingen, Germany
[2] Center for Light-Matter Interaction, Sensors & Analytics LISA+, University of Tübingen, Auf der Morgenstelle 15, D-72076 Tübingen, Germany

**Abstract**

We report a high-throughput and easy-to-implement approach to fabricate microchannels of nanocrystal superlattices with dimensions of ~4 µm$^2$, thus approaching the size of typical single-crystalline domains. By means of microcontact printing, highly ordered superlattices with microscale dimensions are transferred onto photolithographically prepatterned microelectrodes, obtaining well-defined superlattice microchannels. We present step-by-step guidelines for microfabrication, nanocrystal self-assembly and patterning to archive large quantities of up to 330 microchannels per device for statistically meaningful investigations of charge transport in single-crystalline superlattice domains. As proof-of-concept, we perform conductivity and field-effect transistor measurements on microchannels of PbS nanocrystal superlattices. We find that the electric transport within microchannel superlattices is orders of magnitude more efficient than within conventional large-scale channels, highlighting the advantage of the near single-crystalline microchannels presented in this paper.

# 1. Introduction

Colloidal nanocrystals (NCs) are of high interest in academic research and technological applications, due to their size-tuneable optoelectronic properties [1–3]. The self-assembly of NCs into highly ordered superlattices provides a powerful platform for the development of numerous solution-processed optoelectronic devices [4–8]. These periodic arrays mimic classical crystals in which atoms have been replaced by NCs. In analogy to atomic crystals, electronic coupling between neighbouring NCs can occur, resulting in novel collective properties by design. Superlattices of NCs are commonly fabricated by the self-assembly of NCs from solution, comprising the solvent-evaporation based methods of drop-casting, spin-coating, dip-coating and the assembly at the liquid/air interface [9]. This allows NC arrays to be realized with structures raging from 2D monolayers [10–12] to 3D arrays with standard superlattices structures (e.g. bcc, fcc, hcp lattices) [13–15] and more sophisticated binary lattices [16–18]. Further, superlattices with oriented attachment of atomically connected NCs with honeycomb or square lattice structures can be achieved [19–21]. Despite their vast structural versatility, all NC assemblies suffer from one mayor drawback that restricts the exploration of the fundamental electronic properties of these novel artificial solids: the assemblies are rather heterogeneous and typical superlattice grain sizes are yet only a few to tens µm wide [22–25] such that the coupling between NCs is limited to small domains, caused by grain boundaries, cracks, voids and disorder [2, 26, 27]. However, typical devices to investigate the electronic properties of NC superlattices involve active areas with much larger dimensions of ~$10^4$–$10^5$ µm$^2$ [8, 12, 21, 28–30]. Thus, measuring the electronic properties of these superlattice devices averages over the intrinsic properties of all present domains, concealing the expected collective effects of single-domain NC superlattices. Accordingly, to really pinpoint the intrinsic properties of NC superlattices, the active channel area between two contacts has to be decreased to allow single-grain measurements [9, 21, 31]. In this respect, we explicitly do not refer to contactless measurement techniques (e.g. terahertz spectroscopy [12, 27, 32, 33]), since an actual implementation of superlattices into real devices requires contacting by electrodes. First indications of the effect of reducing the measuring dimensions have been provided by Mentzel *et al* [34], where nanopatterned NC films prepared by electron beam lithography have revealed 180× higher conductivities compared to macroscopic films. Evers *et al* [32] have used scanning tunnelling microscope tips with distances of 500 nm to probe field-effect mobilities of single-grain superlattices, reporting that the absence of long-range order is limiting the mobility.

Here, we demonstrate our high-throughput, low-cost and easy-to-implement approach of fabricating microchannels of lead sulfide (PbS) NC superlattices to define active areas with dimensions close to single domains. By means of microcontact printing (μCP), we transfer periodic stripes of highly ordered NC superlattices with microscale dimensions onto photolithographically prepatterned electrode devices. This leaves most of the substrate area uncoated and is key to obtain well-defined microchannels. In the soft-lithography technique of μCP, a patterned elastomeric stamp is used to transfer an NC superlattices to a substrate while the stamp pattern is maintained [35, 36]. Channels with dimensions of ~1 × 4 µm$^2$ fabricated in this way allow performing electronic measurements of single crystalline superlattice domains. We provide a step-by-step protocol to fabricate devices with several hundreds of microchannels for a statistically meaningful investigation of the electronic properties of single-crystalline superlattices. This protocol is not limited to PbS NCs and applicable for a broad range of NC materials and superlattice types.

## 2. Methods

### 2.1. Materials.

All required materials and equipment for the fabrication of microchannels are listed below. Positive tone resist (ma-P 1205), negative tone resist (ma-N 405), developer (ma-D 331/S) and remover (mr-Rem 660) were purchased from micro resist technology GmbH, Polydimethylsiloxane (PDMS) Sylgard 184 prepolymer and cross-linker from Dow Corning GmbH and hexamethyldisilazane (HMDS) as well as Tridecafluoro-(1,1,2,2)-tetrahydrooctyl-trichlorosilane (F$_{13}$TCS) from Sigma Aldrich. For the electrode devices, we used (100)-Si wafers with 200 nm SiO$_2$ (Siegert Wafer GmbH). However, other substrates or wafers can also be used. (100)-Si wafers were used for the stamp master fabrication, ideally with a SiO$_2$ layer of 100–200 nm. Standard chemicals as acetone, isopropanol, ultrapure water, potassium hydroxide (KOH) pellets and buffered oxide etch solution (BOE 7:1, (87.5% NH$_4$F : 12.5% HF)) were used.

Oleic acid stabilized PbS NCs were synthesized according to Weidman *et al* [37]. Cu-4,4',4'',4'''-tetraaminophthalocyanine (Cu4APc) was synthesized according to Jung *et al* [38]. Hexane, octane, acetonitrile and dimethyl sulfoxide, were purchased from Acros Organics.

## 2.2. Fabrication process of Au electrode devices with µm-gaps.

(100)-Si wafers with a 200 nm SiO$_2$ layer were cut into 15 × 15 mm$^2$ pieces. HMDS as an adhesion layer was applied: The samples were heated to 150 °C in a closed glass Petri dish under nitrogen flow. After 15 min, the temperature was set to 120 °C and one drop of HMDS was placed with a syringe next to the wafer pieces while the petri dish was kept closed. After 5 min, the samples were slowly cooled down to room temperature. A thin layer of the negative tone photoresist ma-N 405 was applied (100 µl, maximum spin speed of 10000 rpm, 30s), followed by a soft bake at 95 °C for 60 s. The thickness of the resist was 310 nm, measured by profilometry (Bruker, Dektak XT). For exposure, an optical mask with the electrode structures as opaque parts was used. The substrates were exposed for 45s (365 nm, 325 W, Karl Süss MA6 mask aligner) followed by development for 5–7.5 min in ma-D 331/S to remove the unexposed parts, stopped by placing the sample into ultrapure water. The samples were placed in a thermal evaporation system (Pfeiffer Vacuum PLS 570) under high vacuum conditions (10$^{-7}$ mbar). As an adhesion layer, ~2.5 nm of Ti was evaporated, followed by ~8 nm of Au. The thickness and evaporation rate were controlled by a quartz crystal microbalance. Finally, lift-off was performed in mr-Rem 660 assisted by ultrasonication.

## 2.3. Fabrication of silicon stamp masters.

First, the (100)-Si wafers were thermally oxidized for 75 min at 1050 °C to yield a SiO$_2$ layer of around 100 nm. Alternatively, (100)-Si wafers with an initial SiO$_2$ layer can be used. The Si/SiO$_2$ wafer were cut in 15 × 15 mm$^2$ pieces, with the cutting line precisely aligned to the <110> Si direction (indicated by the wafer flat). HMDS was applied as an adhesion layer, as described above. A 500 nm thick layer of the positive tone photoresist ma-P 1205 was applied (100 µl, 3000 rpm, 30s), followed by a soft bake at 90 °C for 60 s. An optical mask with 15 µm wide stripes as transparent parts was used. The stripes were aligned parallel to the <110> direction. Exposure for 1 s and development for 60 s in ma-D 331/S removes the exposed parts (stripe profile). Next, SiO$_2$ etching was performed in a reactive ion etching (RIE) system (Oxford Instruments, Plasmalab 80 Plus) with the following process parameters: a mixture of CHF$_3$ and O$_2$ (45 sccm and 5 sccm, respectively), a chamber pressure of 40 mTorr and a power of 150 W were used. The ideal etch time was 180 s (see figure 3). Before and after the CHF$_3$-RIE, the substrates were cleaned with an oxygen plasma (50 sccm O$_2$, 10 s, 100 mTorr, 40 W). Residual resist was removed by acetone. Before KOH etching, the substrates were immersed in 1.2% HF solution for 60 s (1 ml of BOE 7:1 in 10 ml ultrapure water) and then directly mounted in a home-built Teflon holder placed in an 8.8 M (36 wt%) KOH solution (27 g KOH in 48 ml ultrapure water) heated to 60 °C via a water bath under stirring. The substrates were etched for 20–40 min and rinsed with ultrapure water to stop the etching process. The substrates were immersed into an 1.2% HF bath for 20 min and rinsed with ultrapure water to remove the $SiO_2$ etch mask. For the surface functionalization, the substrates were heated to 150 °C in a closed Petri dish under nitrogen flow. After 15 min, one drop of $F_{13}TCS$ was placed with a syringe in the Petri dish, which was then kept closed for 30 min. The functionalized silicon masters were slowly cooled down to room temperature and rinsed with acetone and isopropanol to remove excess $F_{13}TCS$.

**2.4. Fabrication of elastomeric stamps.**

The functionalized silicon masters were placed into a home-built Teflon chamber (3 × 3 masters). PDMS base and curing agent (Sylgard 184) were mixed in a 10:1 ratio (33 ml in total to obtain 9 stamps of roughly 10 mm thickness) and stirred thoroughly for ~3 min. The mixture was placed in a vacuum desiccator to remove trapped bubbles by multiple evacuation and re-pressurisation steps and final evacuation for ~20 min. The degassed PDMS mixture was poured onto the silicon masters in the Teflon well. The well was placed in the desiccator and evacuated for another ~5 min to remove air bubbles trapped at the master-PDMS interface. The PDMS was then cured overnight (~16 h) in an oven at 150 °C. Shorter curing times increase the softness of the PDMS stamps. The cured PDMS block was cooled down to room temperature and carefully released from the Teflon well. Since a small amount of PDMS normally also wetted the backside of the silicon master, this thin layer was removed with a razor blade and the masters were then carefully peeled off (parallel to the trenches) from the PDMS block. The stamps were cut into $10 \times 10 \times 10$ mm$^3$ pieces with a razor blade, cleaned by sonication in isopropanol and dried under pressurised nitrogen flow. The cubic shape allows easy handling of the stamps by hand.

**2.5. Self-assembly of nanocrystal superlattice films at the liquid/air interface.**

PbS NCs with a diameter of $5.8 \pm 0.5$ nm, stabilized with oleic acid, were dispersed in hexane:octane with ratios from 1:0–0:1 at concentrations of 4–10 µM. A home-built Teflon chamber with an area of $10 \times 10$ mm$^2$ was filled with 1 ml acetonitrile and equipped with a lid to seal the Teflon chamber. A needle containing a PbS NC dispersion was mounted just above the acetonitrile subphase and connected to a syringe pump. Another needle containing the ligand solution (Cu4APc in dimethyl sulfoxide), connected to another syringe pump, was mounted within the subphase. The PbS NC dispersion (70–100 µl) was injected on top of the acetonitrile subphase, whereas the injection speed was controlled by the syringe pump. The ligand solution (150 µl, ~0.1 mg ml$^{-1}$) was injected at the bottom of the liquid subphase and ligand exchange was performed over a duration of 4 h.

**2.6. Microcontact printing of NC superlattice microchannels.**

A micropatterned PDMS stamp was parallelly brought into contact with the floating NC superlattice film for ~5 s and excess liquid was removed from the stamp with a tissue. The coated stamp was placed onto the Si/SiO$_2$ substrate with prepatterned electrodes for ~30 s and then removed in a tilted manner (parallel to the stripe profile). The stamped substrates were vacuum-dried for a few minutes to remove excess liquid. The substrates were then placed on a spin coater and covered with acetone to remove unbound ligands. After 30 s, the acetone meniscus was removed by spinning at 1200 rpm for 30 s. The washing step was repeated once. This assembly and µCP process was performed in a glovebox in nitrogen atmosphere (level of O$_2$ < 0.5 ppm and H$_2$O = 0 ppm).

**2.7. Characterization of superlattice stripes and microchannels.**

Scanning electron microscopy (SEM) imaging was performed with a HITACHI model SU8030 at 30 kV. For sideview investigation of superlattice stripes, devices were analysed under a tilt angle of 85°. The thickness of the transferred superlattice stripes were investigated by atomic force microscopy (AFM), conducted with a Bruker MultiMode 8-HR. Electrical measurements were performed at room temperature in a nitrogen-flushed probe station (Lake Shore, CRX-6.5K). Individual electrode pairs with a connected superlattice stripe were contacted with W-tips, connected to a source-meter-unit (Keithley, 2636B). A base plate served as third electrode contacting the gate of Si/SiO$_2$ devices. Two-point conductivity measurements of individual microchannels were performed by applying several voltage sweeps of ±1 V and ±200 mV and detecting the current between the electrodes as well as the leak current. Field-effect transistor (FET) measurements were conducted by applying a constant source-drain voltage of $V_{S-D}$ = 5 V, while the current flow was modulated by applying voltage sweeps in the range of -40 V ≤ $V_G$ ≤ 40 V on the gate electrode. The detected current was corrected as $I_{S-D} = I_{S-D}$ (5 $V_{S-D}$) - $I_{S-D}$ (0 $V_{S-D}$) and $I_G = I_G$(5 $V_{S-D}$) – $I_G$ (0 $V_{S-D}$).

The gradual channel approximation was used to calculate the field effect mobility $\mu$ of individual microchannels (see supplementary materials for details).

## 3. Results

We emphasize that the described process can easily be implemented in any laboratory as long as the following standard microfabrication equipment is generally available: Photolithography equipment like a spin coater, a mask aligner and an optical microscope, a thermal evaporation system, a RIE system with $O_2$ and $CHF_3$ as process gases, and other standard laboratory equipment such as Teflon petri dishes and a magnetic hotplate stirrer.

### 3.1. Microchannel device layout.

First, for a µCP process with a high success rate, an elastomeric stamp with an optimal aspect ratio of its relief features has to be designed. Here, the width $W$ of the stripes is set to ~4 µm which is a typical grain size of superlattices. The stamp feature height $H$ and feature distance $D$ are chosen to have an ideal aspect ratio ($H/W$) according to $0.5 < H/W > 5$ and $H/D > 0.05$ to prevent lateral collapse and pairing of the elastomeric stamp features [39–41]. Accordingly, a feature height of $H = 8$ µm and a feature periodicity of 80 µm ($D = 76$ µm) is selected.

Next, the layout of the finger electrodes is adjusted: The electrodes from opposite directions overlap at their ends. The length of the overlap corresponds to the periodicity of the stamp features of 80 µm. Thus, only one orthogonally transferred stripe connects adjacent electrodes. With this device geometry, well-defined microchannels can be realized where an entirely homogeneous electric field is established within the channel and the direction of the electric field vector is well-known. One set of 12 electrodes form 11 individually addressable microchannels. On one substrate, 30 of these electrode sets are present, yielding a total number of 330 microchannels per device (see figure S1 in the supplementary material).

### 3.2. Fabrication process of Au electrode devices with µm-gaps.

To warrant high throughput, Au electrodes with ~1 µm gaps are fabricated by UV photolithography (figure 1). The lift-off technique with a negative tone resist is chosen for high-resolution patterning. As photolithography is diffraction limited, the resolutions scales with ~$\sqrt{\lambda h}$, where $\lambda$ is the wavelength of exposure and $h$ the thickness of the photoresist [42]. Next to a short exposure wavelength, a resist thickness as thin as possible while retaining its functionality is required. This was empirically identified as a thickness in the range of 250–350 nm, achieved by spin-coating the substrate with a low viscosity resist at maximum spin speed. Furthermore, the photomask exhibits opaque electrode structures which are separated by gaps of only 0.6 µm. The layout of the photomask is given in figure S1. Due to diffraction, the sub-µm gaps result in ~1 µm photolithography pattern transfer. Negative tone resist tends to form undercut structures after development (figure 1b), which are well-suited for lift-off processes. After metallization with ~2.5 nm Ti and ~8 nm Au, the lift-off of the residual resist and metal layer on top results in Au electrodes with ~1 µm gaps, referred to as channel length *L*, and smooth edges (figure 1c). Thin electrodes with smooth edges are desirable as they prevent breaking of the transferred superlattice stripes at the edges.

The presented guideline for the fabricated Au electrodes with ~1 µm gaps on Si wafers with an $SiO_2$ layer of 200 nm allows to perform e.g. two- and four-point measurements as well as field-effect transistor measurements. Generally, this approach can easily be adjusted for other substrates, such as glass slides or polyimide membranes.

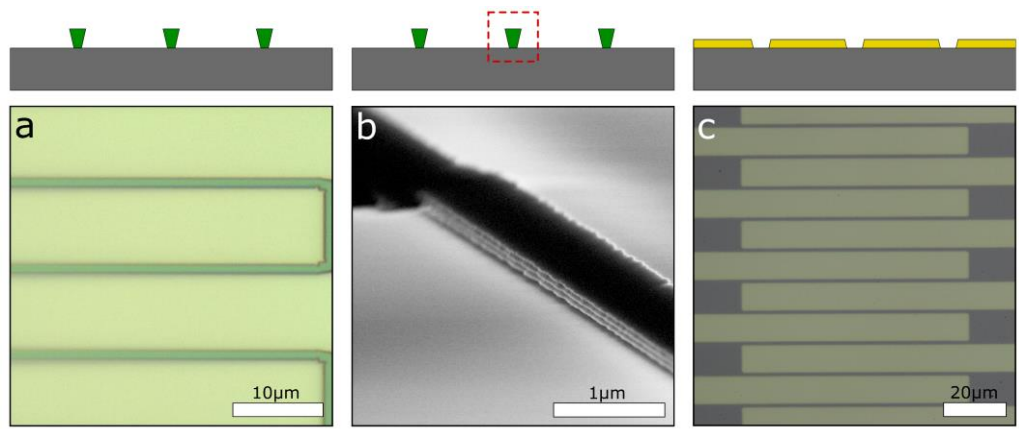

**Figure 1**. Fabrication of prepatterned electrode devices. (a) Optical micrograph of a Si/$SiO_2$ substrate with developed negative tone resist pattern (dark green) of ≤1 µm width and a thickness of 310 nm. (b) SEM micrograph of an ~800 nm wide resist patterns under a tilted view of 85°. (c) Optical micrograph of the substrate after metallization and lift-off. Well-defined electrodes with gaps of ≤1 µm are formed. The respective schematic drawings are provided at the top of each subfigure.

### 3.3. Fabrication process of silicon masters and elastomeric stamps.

To meet the above-mentioned requirements for the elastomeric stamps, especially the relatively large relief height of 8 µm, a silicon master fabrication by means of anisotropic etching in aqueous KOH solution was chosen. We used (100)-Si wafers for the etching process, as this can produce trenches with inclined {111} walls [43]. The etch rate of silicon differs for the crystal planes, caused by different activation energies. The selectivity of the etch rate of the crystal planes {100}:{111} is about 100:1 [44]. Thus, the {111} plane effectively serves as an etch stop. This technique provides the advantages of atomically flat defined sidewalls and the possibility to adjust the width of the stamp features by controlling the etching depth/time, while the inclined (non-vertical) sidewalls of the final elastomeric stamps improve their stability. In comparison, photoresist-based masters suffer from only poorly defined side-walls and lower thermal stability. This may alter the shape of the resist profiles during the fabrication and molding (PDMS baking) process. Further, changing the relief width requires a new lithography mask.

The fabrication process of stamp masters of atomically precise silicon relief with dimensions introduced above is schematically illustrated in figure 2 and detailed in the methods. The relief pattern can be changed considering the standard microcontact printing concepts [39] and the process can easily be adjusted.

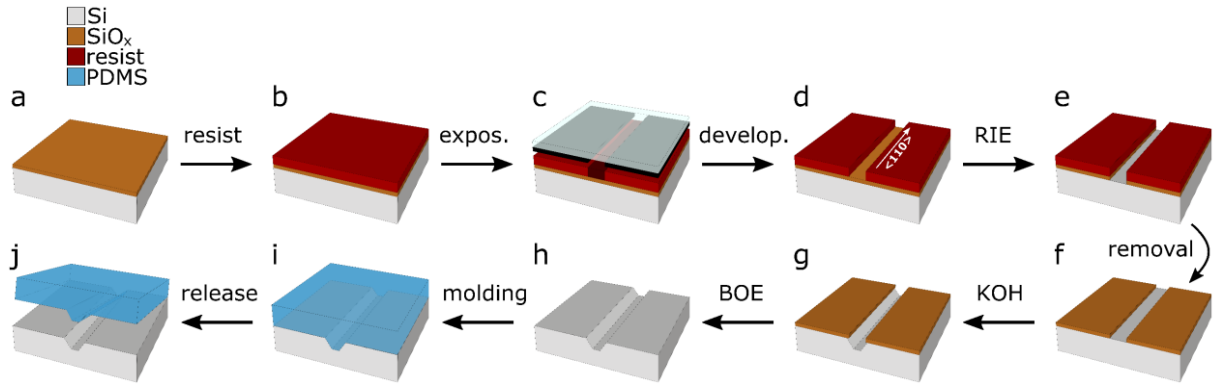

**Figure 2**. Fabrication process of silicon master. (a) (100) Si wafer with a 100 nm thick $SiO_2$ layer. (b) Positive tone resist (~500 nm) is applied, (c) UV exposure with an optical mask with transparent stripes of 15 µm width and a periodicity of 80 µm. The stripes are aligned parallel to the <110> Si direction. (d) After development, $SiO_2$ is exposed and (e) removed via reactive ion etching. (f) Residual resist is removed, exposing the patterned $SiO_2$, which serves as an etch mask in the anisotropic KOH etching (g). The {111} plane serves as etch stop, resulting in inclined trenches with an angle of 54.7°, a base width of ~4 µm and a depth of ~8 µm. (h) Removal of the $SiO_2$ layer with buffered oxide etch yields Si masters. After functionalization with $F_{13}TCS$ as an anti-sticking layer, they can be used as molds for soft-lithography stamps made of PDMS with the negative relief (i,j).

First, the desired stripe profile pattern is transferred by standard photolithography, using positive tone resist with high stability against dry etching processes (figure 2a–d). Alternatively, negative tone resist together with an inverted photomask could be used. The photomask contains a stripe pattern with a periodicity of 80 µm, matching the electrode overlaps. Considering the angle of 54.7° between the {100} and {111} silicon planes, the width of the stripes is set to 15 µm (supplementary material and figure S2 for details). This finally results in 4 µm wide trenches after etching to a depth of 8 µm. Since the desired anisotropic etching of Si strongly depends on the orientation of the crystallographic planes, it is crucial that the patterns are parallel to the <110> direction (figure 2d).

To transfer the pattern into the $SiO_2$, which serves as a mask during KOH etching, the substrates are etched in a RIE system using $CHF_3$ (figure 2e) due to its high selectivity between silicon and $SiO_2$ [45]. The residual resist is removed and the depth of the etched trenches are measured by profilometry (figure 2f). As apparent in figure 3, after etching for 150 s the etch rate declines from ~42 nm min$^{-1}$ for $SiO_2$ to ~2 nm min$^{-1}$ for Si, as the 100 nm thick $SiO_2$ layer is fully removed and the silicon surface is exposed. Thus, an etch time of $t > 150$ s is chosen to ensure a full removal of the exposed $SiO_2$ areas ($t = 180$ s).

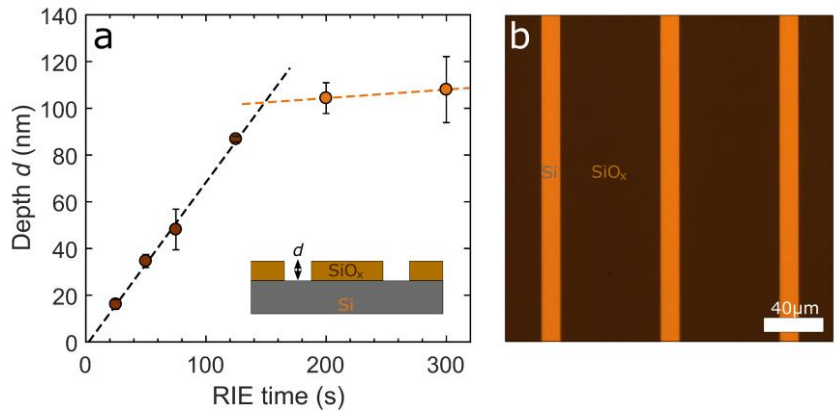

**Figure 3**. Etch rate of $SiO_2$ by $CHF_3$-RIE. (a) Measured depth of the etched part as a function of etch time ($CHF_3$-RIE). After etching for 150 s, the etch rate declines as the $SiO_2$ is fully removed and the Si surface is exposed. The etch rate at $t < 150$ s is ~42 nm min$^{-1}$ and corresponds to $SiO_2$, whereas at $t > 150$ s the rate is ~2 nm min$^{-1}$ and corresponds to Si. The thickness of $SiO_2$ can be determined to be ~100 nm. Error bars represent the standard deviation of several measurements. The colour code indicates the transition from $SiO_2$ to Si. This step corresponds to figure 1d–f. (b) Optical micrograph of a sample after RIE and resist removal. The trenches in the $SiO_2$ layer exposing bare Si can easily be identified. This corresponds to the schematic drawing in figure 2f.

The $SiO_2$ pattern prepared in this way serves as an etch mask during the KOH etching, which allows to transfer the pattern into silicon. Before KOH etching, the native ~2 nm thick $SiO_2$ layer formed at ambient conditions [46] needs to be removed by an HF dip.

During anisotropic KOH etching, the substrates are mounted in a home-built Teflon holder that enables to suspend the substrates vertically into the KOH bath, with the line pattern parallel to the flow direction of the stirred KOH solution (figure 2g). This facilitates the removal of $H_2$ bubbles formed during etching. After 40 min, trenches with base widths of ~4 µm are observed, corresponding to a etch depth of ~8 µm, as shown in figure 4a,b. The line patterns are oriented parallel to the <110> silicon direction, resulting in well-defined trenches with an inclination angle of 54.7°, caused by the {111} facets serving as an etch stop. The undercut at the edges of the $SiO_2$ etch mask can clearly be identified, as well as residual $SiO_2$ agglomerations accumulated at the base of the trench. The undercut of several hundreds of nm is formed as etching of the {111} facet is marginally present. The experimental etch rate of the {100} facet is calculated to be ~200–250 nm min$^{-1}$. This is in good agreement with the expected etch rate of ~300 nm min$^{-1}$ by Seidel *et al* for an 8.8 M KOH solution at 60 °C, which further shows a $SiO_2$ etch rate of ~0.1 nm min$^{-1}$ [44].

To remove the residual SiO$_2$ etch mask and agglomerations at the base of the trenches, the substrates are treated with BOE (figure 2h), resulting in atomically defined side walls (figure 4c–f).

Finally, the surfaces of the patterned Si substrates are functionalized with F$_{13}$TCS as an anti-sticking layer, adapted from Beck *et al* [47], which prevents the sticking of the cured elastomeric stamps to the silicon surface [39, 47]. When showing a highly hydrophobic surface, the modification of the samples was successful. The final silicon masters can be reused dozens of times to serve as molds for PDMS curing (figure 2i,j). Figure 4g,h shows a cured PDMS stamp, which is exactly the negative of the stamp master: a stripe relief with a periodicity of 80 µm, where the individual protrusions have an inclination angle of 54.7° and a base width of ~4 µm.

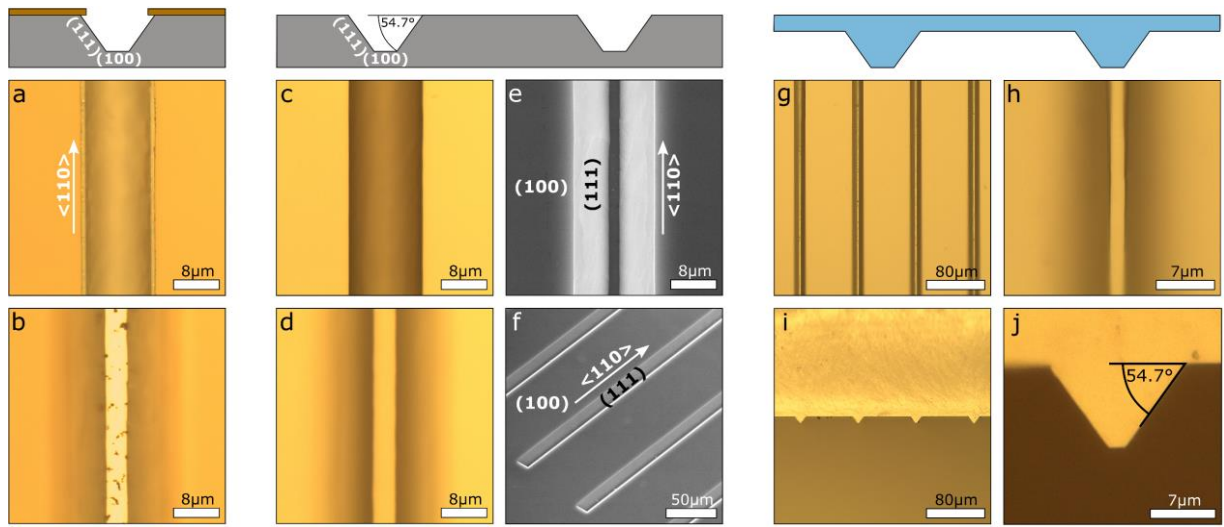

**Figure 4**. Stamp master fabrication and molding of PDMS stamps. (a,b) Optical micrographs of the stamp master after anisotropic KOH etching for 40 min, focused at the upper edge (a) and the base (b), respectively. (c,d) Optical micrographs of the same stamp master after removal of the SiO$_2$ etching mask by BOE treatment. The SiO$_2$ undercut (c) and the residual SiO$_2$ parts accumulated at the base (d) are clearly removed, resulting in atomically defined side walls. (e,f) Corresponding SEM micrographs of a stamp master in top view (e) and under a tilt view of 40° (f). (g,h) Optical micrographs of the cured PDMS stamp, which is exactly the negative of the stamp master, given by the inclination angle of 54.7°. The corresponding side-views of the PDMS stamp are shown in (i,j), respectively. The respective schematic drawings are drawn at the top of each subfigure.

### 3.4. Microcontact printing of microchannels.

We emphasize that the fabrication of superlattice microchannels is neither limited to the nanocrystals/nanoparticles we have used, nor to self-assembly of the NCs at the liquid/air interface.

Here, we use PbS NCs functionalized with the organic π-system Cu4APc as a model system. Cu4APc replaces the native oleic acid ligands, resulting in highly conductive superlattices together with long-range ordered [12]. The organic semiconductor Cu4APc with its functional groups serves as a linker between adjacent NCs and couples them chemically through binding to the NC surface (long range order) and electronically via potentially near-resonant alignment of suitable energy levels and reducing the energy barrier between adjacent NCs (enhanced conductivity) [48, 49].

As the native oleic acid stabilized PbS NCs are soluble in alkanes, the liquid/air interface method developed by Dong *et al* can be applied [28]. This is schematically illustrated in figure 5a–c. The NC dispersion is spread onto the surface of a polar liquid and after evaporation of the solvent, the NCs self-assemble into a floating superlattice film. The thickness and homogeneity of the film can be controlled to a certain degree (e.g. monolayer vs. thick film) by changing the NC dispersion volume, concentration, injection speed, and the evaporation controlled by the adjustable sealing. After injecting a ligand solution into the polar subphase, the Cu4APc ligands diffuse through the liquid subphase to the NC superlattice film and replace the insulating native oleic acid ligands, which renders the highly ordered superlattice conductive [12]. The floating superlattice film can now be transferred onto a patterned PDMS stamp by approaching the stamp onto the liquid/air interface (figure 5d,e). Well-defined stripes of the NC superlattice film are finally printed to a solid substrate with prepatterned electrodes by microcontact printing, as displayed in figure 5f,g. Adjacent electrodes are connected by an orthogonal superlattice stripe, forming an individual microchannels with length $L \approx 1$ µm and width $W \approx 4$ µm, respectively (figure 5h).

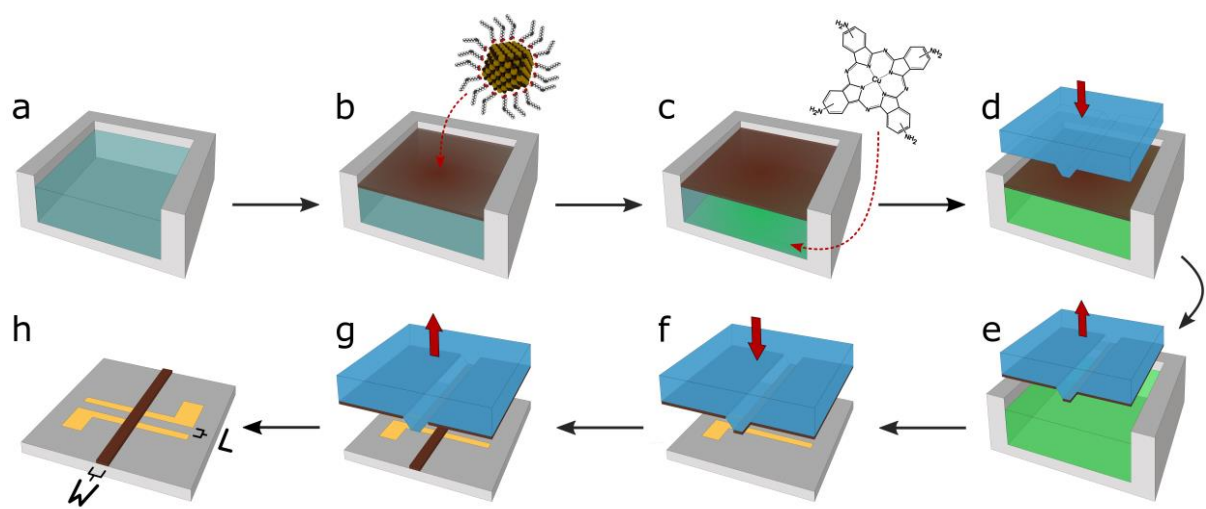

**Figure 5**. Schematic of the self-assembly and ligand exchange of PbS NC superlattices and microcontact printing to form microchannels. (a) A home-built Teflon chamber is filled with acetonitrile as the liquid subphase. (b) A dispersion of oleic acid stabilized PbS NCs is injected on top of the subphase and as the dispersion solvent evaporates, the NCs form a freely floating superlattice film. (c) The ligand solution

(Cu4APc in dimethyl sulfoxide) is injected into the bottom of the liquid subphase, the Cu4APc ligands diffuse through the subphase and replace the native oleic acid ligands of the PbS NCs. (d,e) A micropatterned PDMS stamp is inked with the superlattice film by parallelly approaching the floating film. (f,g) The inked stamp is brought in contact with the electrode device, transferring stripes of the self-assembled superlattice film. (h) Adjacent electrodes are connected by an orthogonal superlattice stripe, forming an individual microchannels with length $L \approx 1$ µm and width $W \approx 4$ µm, respectively. Up to 330 microchannels can be realised per device.

Figure 6a–d shows a PDMS stamp coated with a superlattice film before and after microcontact printing. After stamping onto a substrate, the film is successfully transferred from the base of the stripes, whereas the spaces between the bases are still coated. Accordingly, µm-patterned areas of superlattice films can be transferred (figure 6e). The own weight of the ~1 cm$^3$ sized PDMS stamp is sufficient for a highly successful transfer rate of the superlattice films. The application of additional pressure deforms the elastomeric PDMS stamp (sagging) and, consequently, parts of the interspace between the protruding stripes are transferred, which is undesired (figure 6f,g).

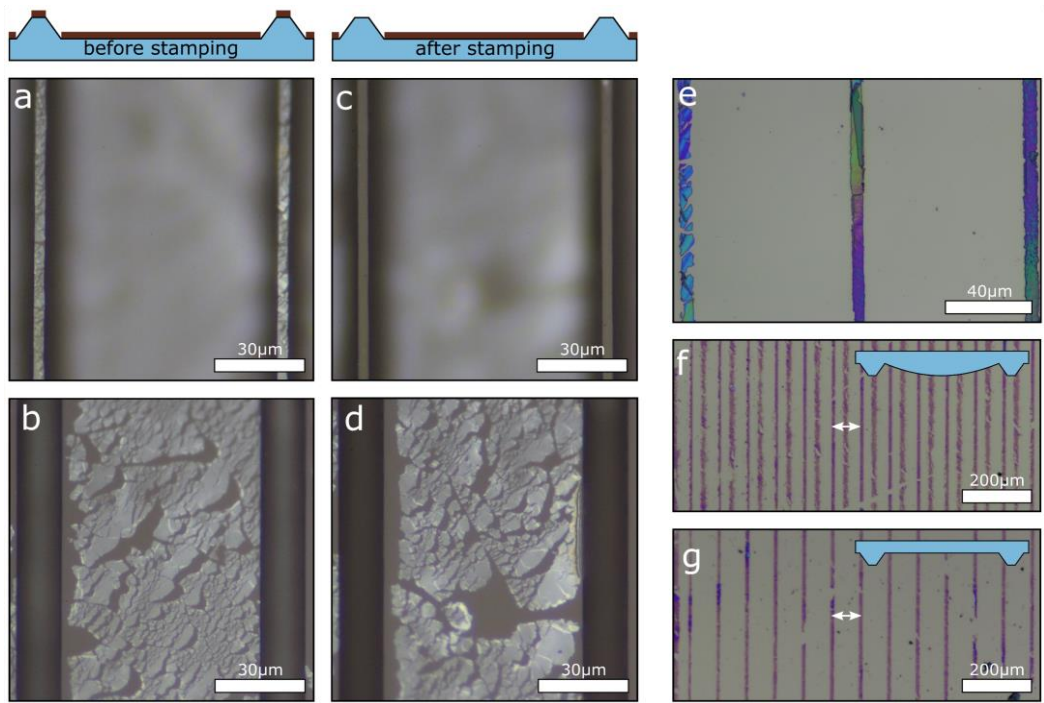

**Figure 6**. Soft-lithography stamps coated with NC thin films. (a,b) Optical micrographs of a PDMS stamp after coating with a NC superlattice film. Both the base of the stripe (a) as well as the space between the stripes (b) are coated. (c,d) After stamping the PDMS stamp onto a substrate, the superlattice film is transferred from the stripe base (c), whereas the space between the bases is still covered (d). The respective schematic drawings are drawn at the top of each subfigure. (e) Optical micrograph of three microcontact printed superlattice stripes on a Si/SiO$_2$ substrate. The different colorations indicate the need for microchannels to characterize single-crystalline domains. (f,g) Optical micrograph of a Si/SiO$_2$ substrate after microcontact printing with a coated PDMS stamp. (f) Pressure which is to high causes deformation of the elastomeric stamp and parts of the space between the protruding stripes are transferred. (g) The own weight of the PDMS stamp is sufficient for a transfer of the thin film with ~100% success rate. The white arrows indicate the periodicity of 80 µm.

Figure 7a,b displays microchannels of self-assembled NC superlattices, where one of the transferred well-defined superlattice stripe connects adjacent electrodes. The formed microchannel is defined by the length between two adjacent electrodes ($L \approx 1$ µm), the width of the transferred stripe ($W \approx 4$ µm) and the thickness $h$ of the NC superlattice (figure 7c,d). Remarkably, superlattices of different thicknesses (ranging from continuous monolayer to 3D films with thicknesses of up to 2 µm) can be transferred (figure 7e), as the film morphology (meaning areas of different thicknesses) is preserved during the stripe stamping process.

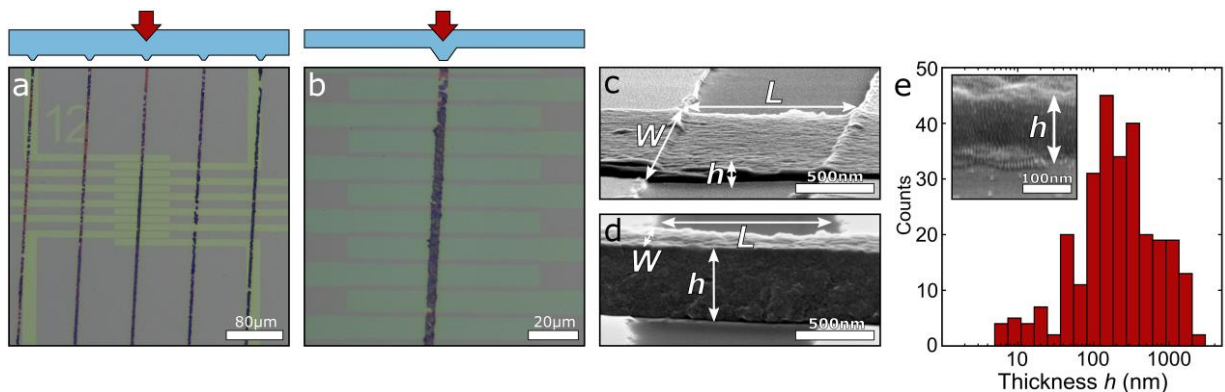

**Figure 7**. PbS NC superlattice microchannels formed by microcontact printing. (a,b) Optical micrographs of a set of electrode pairs (yellow) on a Si/SiO$_2$ device, where an orthogonally stamped PbS NC stripe (brown) connects adjacent electrodes to form microchannels which can be contacted individually. The schematics indicate the microcontact printing process with the patterned stamp. (c,d) Scanning electron micrographs in sideview (85° from normal) of typical microchannels of PbS NC superlattice stripes microcontact-printed across two Au electrodes. (e) Distribution of the thickness $h$ of transferred superlattice stripes shows that superlattices in a wide range of thicknesses can be transferred ($n = 276$ from several samples). Inset: High-resolution scanning electron micrograph of the well-defined edge of a transferred NC superlattice stripe (sample tilted by 85° from normal).

### 3.5. Electronic measurements of superlattice microchannels.

Figure 8a shows a microchannel, where the NCs within the superlattice are highly ordered over the entire channel dimension (figure S3). In these microchannels, the dimensions of the channel itself and the grain-size of typical superlattices match. Remarkably, the structural order of the micrometer-sized crystalline superlattice domains is preserved, which was not achieved in previous attempts [50]. After successful fabrication of superlattice microchannels, a proof of concept of electronic measurements is given in the following.

Figure 8b displays a characteristic two-point conductance measurement of one individual microchannel, showing Ohmic behaviour in the range of ±1 V. This allows determining the conductance value $G$, which is the slope of the $I$-$V$ curve, and the conductivity $\sigma = (G \times L)/(W \times h)$. Measuring approximately two hundred individual microchannels yields a distribution of electric conductivities in the range of $\sigma = 10^{-6}$–$10^{-3}$ S/m.

In addition to conductivity measurements, field-effect transistor measurements can be performed on individual microchannels, as shown in the exemplary transconductance curve in figure 8c. The microchannels consisting of PbS NCs functionalized with the organic π-system Cu4APc show p-type behaviour, which agrees with our previous study on the same material [12]. Using the gradual channel approximation (see supplementary material for details), the field-effect hole mobilities $\mu(h^+)$ can be estimated to be in the range of $\mu(h^+) = 10^{-6}$–$10^{-4}$ cm$^2$ V$^{-1}$ s$^{-1}$, reaching values of up to $2 \times 10^{-4}$ cm$^2$ V$^{-1}$ s$^{-1}$ (figure 8d). The transconductance curve exhibit a hysteresis, which is commonly observed in NC transistors and attributed to charge carrier trapping at the dielectric/NC interface or NC trap states [51].

Finally, we compare the tailored microchannels with conventional channels, where interdigitated electrodes probe large areas of ~1–20 × 10$^4$ µm$^2$ (with 2.5 µm ≤ $L$ ≤ 50 µm and $W$ up to 1 cm), as displayed in figure 8e. Accordingly, using such conventional channel averages over different superlattice domains and their properties present in the channel. Figure 8f shows the geometry-normalized conductance values $G_{\text{geom}} = G \times (L/W)$ of conventional channels and microchannels. For large conventional electrode devices with inhomogeneous superlattice coatings, the conductivity cannot be calculated due to nonuniform thicknesses. However, the distributions of $G_{\text{geom}}$ can clearly be separated, and it is apparent that the normalized conductance of microchannels exceed that of conventional interdigitated electrode devices. Thus, electric transport within microchannel superlattices is orders of magnitude more efficient than within the larger state-of-the-art electrode system. This highlights the advantage of the near single-crystalline microchannels presented in this paper.

However, some limitations should be noted. Here, the superlattice film morphology can only be controlled to a certain degree. As the morphology is preserved during the stripe stamping process, microchannels with different superlattice thicknesses are obtained. Further optimizing the liquid/air interface method could yield more homogeneous thickness distributions.

In addition, to the effect of single-crystallinity, the effect of an increased probability for the formation of percolative networks might play a role.

In a previous report, we have already verified that contact resistance of these materials on the conventional substrates is insignificant compared to the bulk resistance [12]. While we cannot rule out that contact resistance may play a role with the microchannels presented here, its effect would most probably be an increase of the overall resistance. Thus, it cannot explain the improved transport properties shown here in figure 8.

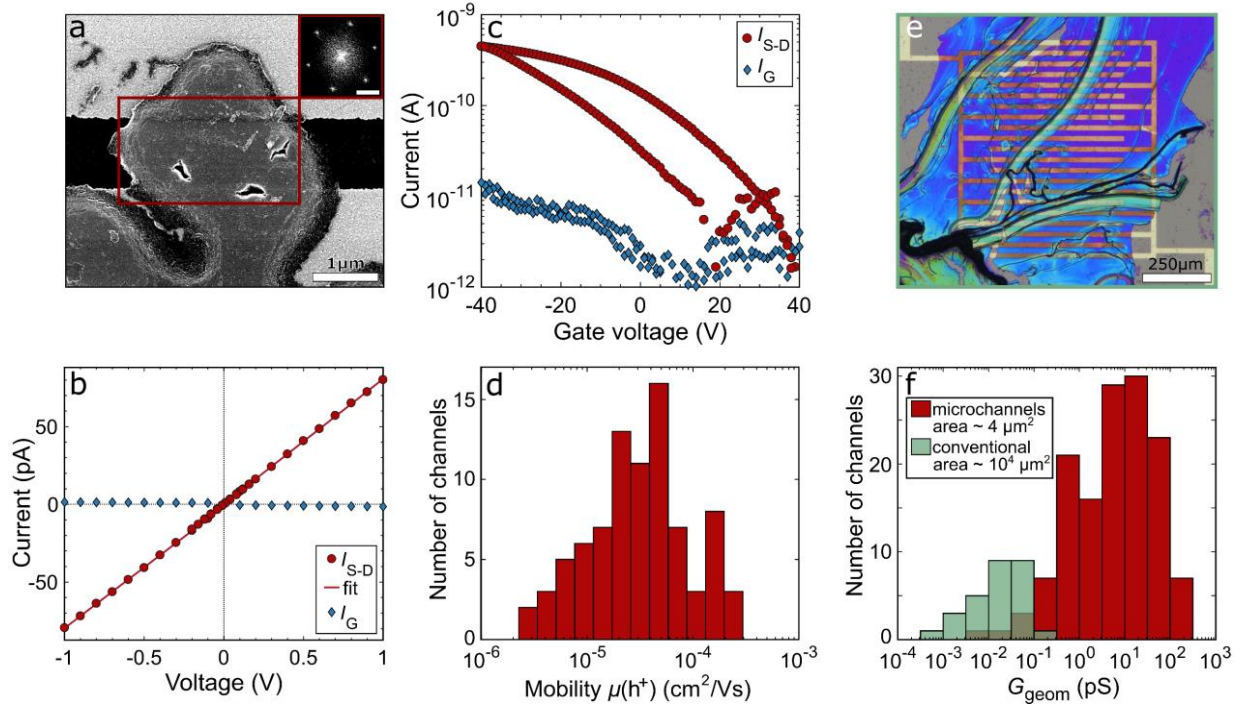

**Figure 8.** Electronic measurements of superlattice microchannels. (a) Scanning electron micrograph of a typical superlattice microchannel with $L = 1$ µm and $W = 3.0 \pm 0.4$ µm. The inset shows the fast Fourier transform, taken at the highlighted area (red box), indicating highly-ordered NCs over the entire microchannel. The scale bar of the inset corresponds to 5 nm$^{-1}$. A high-resolution micrograph of this superlattice is given in figure S3. (b) Typical $I$-$V$-curve of a microchannel of a PbS NC superlattice, showing Ohmic behaviour. The linear fit of the current $I_{S-D}$ yields the conductance. The leak current $I_G$ is negligible ($V_G = 0$ V). (c) Transconductance curve of a microchannel where the source-drain current $I_{S-D}$ is modulated by the applied gate voltage $V_G$ (constant source-drain voltage of $V_{S-D} = 5$ V). The leak current $I_G$ is orders of magnitude lower. (b) Distribution of field-effect hole mobilities $\mu(h^+)$ of $n = 84$ individual microchannels. (e) Optical micrograph of a typical conventional channel coated with a PbS NC superlattice film. Interdigitated electrodes probe areas of ~$10^4$ µm$^2$ ($L = 20$ µm, $W = 1$ cm). (f) Distribution of geometry-normalized conductance of conventional channels (green, $n = 28$) and microchannels (red, $n = 211$), $V_G = 0$ V. Here, $G_{geom} = G \times (L/W)$.

## 4. Conclusion

To conclude, we have fabricated microchannels of NC superlattices, where the channel dimensions approach the typical grain size of self-assembled superlattices of a few µm$^2$ by using the soft-lithography technique of microcontact printing. This was achieved by combining the top-down processes of anisotropic etching of silicon and photolithography together with the bottom-up process of NC self-assembly and ligand exchange and the liquid/air interface. Step-by-step protocols are provided which are easily adjustable for different NC and superlattice types. We demonstrate proof-of-principle by fabricating microchannels of near single-crystalline domains of highly ordered coupled PbS NCs and measuring the electric conductivities as well as field-effect mobilities. A comparison of the superlattice microchannels with conventional state-of-the-art electrode devices reveals the advantageous effect of the near single-crystalline microchannels, presented in this paper. These microchannel NC superlattices enable novel opportunities for studying fundamental physical properties of NC ensembles, such as anisotropic electric transport [52].

## Acknowledgements

This project has been funded by the European Research Council (ERC) under the European Union's Horizon 2020 research and innovation program (grant agreement No 802822). SEM measurements using a Hitachi SU 8030 SEM were funded by the DFG under contract INST 37/829-1 FUGG. The authors thank Monika Fleischer for providing access to their clean room facility. The authors declare no competing interests.

## Conflict of interest

The authors declare no conflict of interest.

# – Supplementary Material –

# Fabrication of nanocrystal superlattice microchannels by soft-lithography for electronic measurements of single-crystalline domains

**Andre Maier[1,2]\*, Ronny Löffler[2] and Marcus Scheele[1,2]**

[1] Institute of Physical and Theoretical Chemistry, University of Tübingen, Auf der Morgenstelle 18, D-72076 Tübingen, Germany
[2] Center for Light-Matter Interaction, Sensors & Analytics LISA+, University of Tübingen, Auf der Morgenstelle 15, D-72076 Tübingen, Germany

Figure S1 displays the design layout and the photomask for the fabrication process of electrodes with µm-sized gaps. As negative tone resist is used for high-resolution patterning, the electrode structures are opaque.

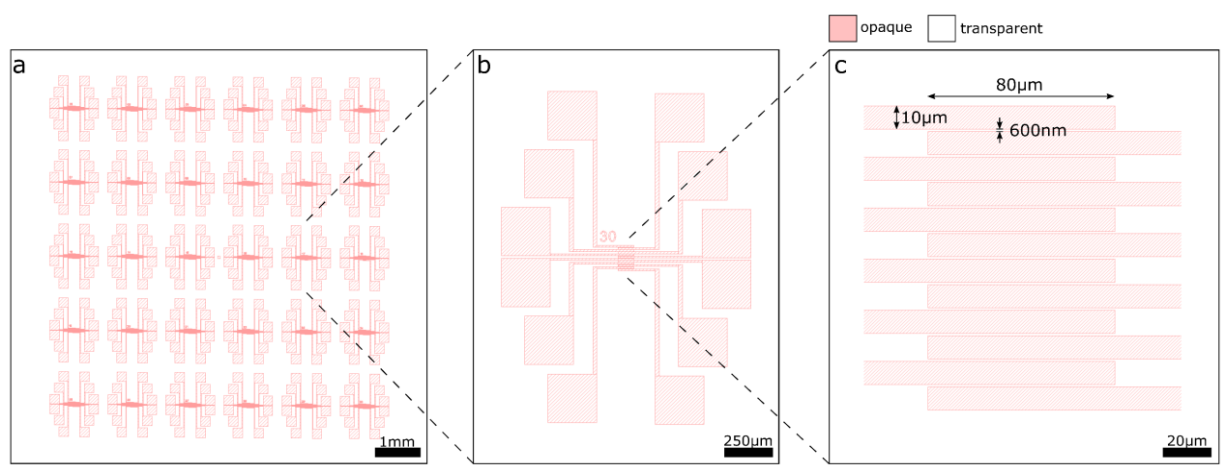

**Figure S1**. Layout of the photomask to create electrode gaps for the microchannels. (a) Schematic of the photomask of an entire substrate with $15 \times 15$ mm$^2$. 30 electrode sets are arranged in a $5 \times 6$ matrix. (b) Schematic of an individual electrode set consisting of 12 electrodes converging from opposed directions in an alternating manner. Each electrode can be contacted by a $250 \times 250$ µm$^2$ contact pad. Labels provide orientation. (c) The alternating electrodes form 11 electrode gaps with an overlap of 80 µm. The distance between two electrodes is set to 600 nm. In total, 330 electrode gaps are present.

Figure S2 displays the design layout and the photomask for the fabrication process of the stamp masters. The stripes are transparent, since positive tone resist was used. As the final stamp dimensions width and height are set to $W = 4$ µm and $H = 8$ µm, respectively, the width of the photomask stripes is $W^* = 15$ µm, considering the angle of 54.7° between the (100) and (111) Si plane: $x = H / \tan(54.7°) = 5.6$ µm and $W^* = 2x + W = 15.2$ µm.

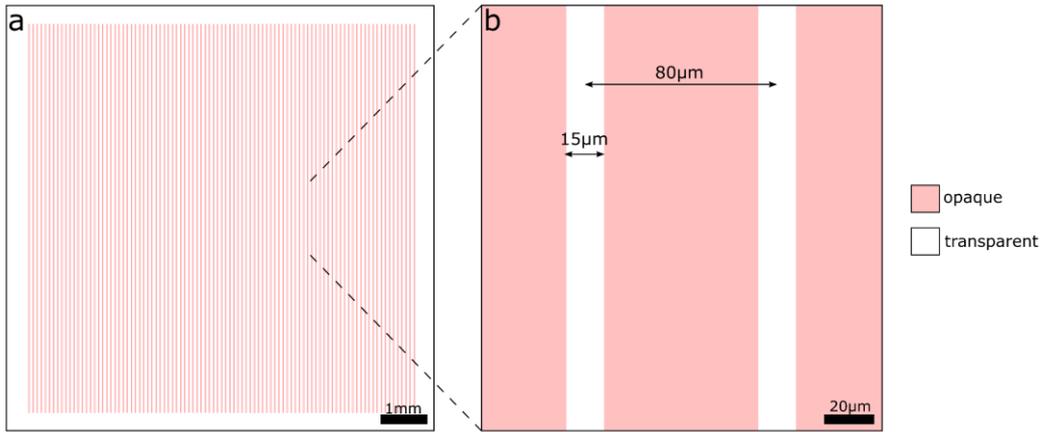

**Figure S2**. Layout of the photomask to fabricate Si-based stamp masters. (a) Schematic of the photomask of an entire 15 × 15 mm² substrate. (b) Transparent stripes with periodicity of 80 µm and a width of 15 µm are present.

To calculate the field-effect mobility µ of individual microchannels, the gradual channel equation is used, given in equation S1 [1].

$$\mu = \frac{\partial I_{S-D}}{\partial V_G} \frac{L}{W} \frac{t_{ox}}{\varepsilon_0 \varepsilon_r V_{S-D}} \qquad (S1)$$

$\frac{\partial I_{S-D}}{\partial V_G}$ corresponds to the derivation of the transconductance curve ($I_{S-D}$ as the detected source-drain current and $V_G$ as the applied gate voltage). $L$ and $W$ are the microchannels length and width. $t_{ox}$ and $\varepsilon_0 \varepsilon_r$ are the thickness and the permittivity of the dielectric $SiO_x$ layer, respectively. $V_{S-D}$ corresponds to the applied source-drain voltage. While the geometry of our microchannels is not ideal for FET measurements, this approach is sufficient for a qualitative comparison of different microchannels.

Scanning electron micrographs of transferred PbS NC superlattices are given in figure S3.

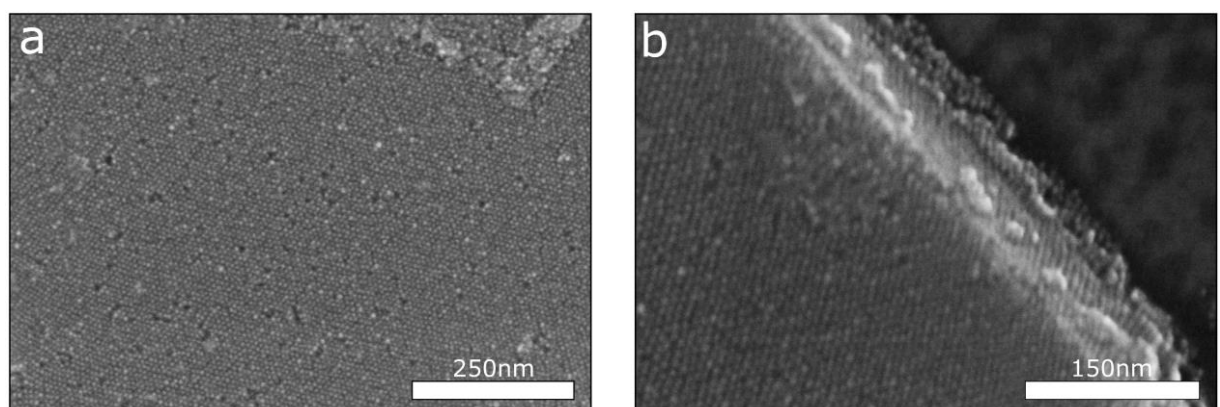

**Figure S3**. Scanning electron micrograph of typical PbS NC superlattices transferred via microcontact printing. (a) Magnification of the superlattice of the microchannel shown in figure 8a. (b). High-resolution micrograph of a superlattice edge. Well-defined edges are present and structural order is preserved.

An atomic force micrograph of a superlattice microchannel is given in figure S4.

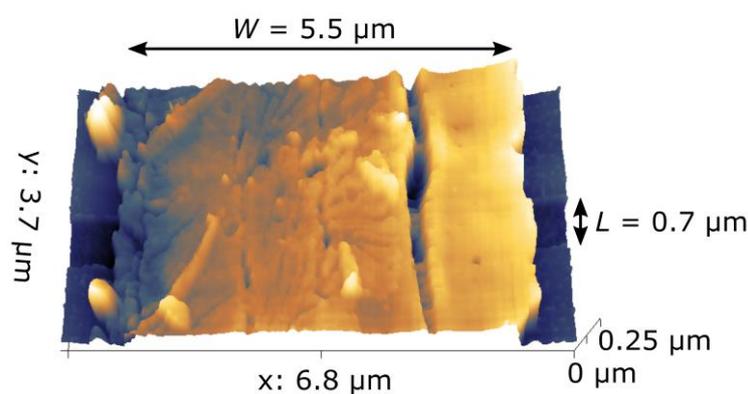

**Figure S4**. Atomic force micrograph of a typical superlattice microchannel. The dimensions of the microchannel are $L = 0.7$ µm, $W = 5.5$ µm and $h = 250$ nm. Additional to the 4 µm wide stamped superlattice (left side), an additional part from the inclined side-wall is transferred (right side).

## Supplementary References